\documentclass[3p,times]{elsarticle}

\usepackage{ecrc}

\volume{00}

\firstpage{1}

\journalname{Physics Procedia}

\runauth{S. Gardner and D. He}

\jid{phpro}

\jnltitlelogo{Physics Procedia}

\usepackage{amssymb}
\usepackage[figuresright]{rotating}

\begin{document}

\begin{frontmatter}



\dochead{}

\title{Tracing Remnants of the Baryon Vector Current Anomaly in Neutron Radiative $\beta$-Decay}


\author{Susan Gardner and Daheng He}

\address{Department of Physics and Astronomy, University of Kentucky, 
Lexington, KY 40506-0055 USA}

\begin{abstract}
We show that a triple-product correlation in the neutron radiative
$\beta$-decay rate, characterized by the kinematical variable
$\eta\equiv (\mathbf{l}_e\times\mathbf{l}_\nu)\cdot\mathbf{k}$, 
where $n(p) \rightarrow p(p') + e^-(l_e) + \overline{\nu}_e(l_\nu) + \gamma(k)$, 
isolates the pseudo-Chern-Simons term found by 
Harvey, Hill, and Hill as a consequence of the baryon vector current
anomaly and SU(2)$_{L}$$\times$U(1)$_Y$ gauge invariance at low energies. 
The correlation appears if the imaginary part of the coupling constant
is nonzero, so that its observation at anticipated levels of sensitivity would reflect the
presence of sources of CP violation beyond the Standard Model. 
We compute the size of the asymmetry in $n\to p e^- \bar\nu_e \gamma$ decay 
as a function of the coupling, estimate the effect of 
Standard-Model final-state interactions, and discuss the
role nuclear processes can play in discovering the effect. 
\end{abstract}

\begin{keyword}
$\beta$-decay \sep decay correlations \sep CP violation 

%% PACS codes here, in the form: 
\PACS 23.40.-s \sep 11.30.Er \sep 12.39.Fe \sep 12.15.-y


\end{keyword}

\end{frontmatter}

% \linenumbers

\section{Introduction}
\label{intro}

Harvey, Hill, and Hill have found that unexpected interactions 
involving the nucleon $N$, photon $\gamma$,  and weak gauge bosons at low energies 
emerge from gauging the axial anomaly of QCD under the full electroweak symmetry 
of the Standard Model (SM)~\cite{HHH2007,HHH2008}. In this contribution we consider how 
such interactions can be isolated through a triple-product momentum
correlation in neutron radiative $\beta$-decay. The correlation is both
parity- and naively time-reversal-odd, so that it vanishes in the Standard Model
save for effects induced by final-state interactions (FSI). Nevertheless, the correlation can 
be generated by sources of CP violation 
beyond the Standard Model, and such couplings, being spin-independent, 
are not constrained by 
the nonobservation of the neutron electric dipole moment (EDM). 
We consider the sorts of limits on its strength which can be determined at existing
and anticipated facilities, as well 
as the size of induced correlations from known FSI.
We also briefly consider the possibility of 
nuclear $\beta$-decay studies as well as 
the prospects for muon-induced reaction studies.

\section{Anomalous interactions at low energies} 
\label{HHH}

The study of the low-energy spectrum of quantum chromodynamics (QCD) 
reveals light, pseudoscalar mesons which we 
interpret as the Nambu-Goldstone bosons of a spontaneously broken chiral symmetry, 
so that in two-flavor QCD, e.g., 
SU(2)$_L\times$SU(2)$_R \rightarrow$ SU(2)$_V$~\cite{nambu,goldstone}. 
A chiral theory of mesons
built on this symmetry is incomplete; the explicit inclusion of the axial anomaly, 
vis-a-vis the Wess-Zumino-Witten (WZW) term~\cite{WZW,witten}, is required in order to describe 
observed processes such as 
$K\bar K \to 3\pi$ and $\pi^0 \to \gamma\gamma$~\cite{witten83}. The latter reflects the
nonconservation of the axial current and hence the presence of the 
axial anomaly~\cite{belljackiw1969,adler1969}. 
If we study the gauge invariance of the WZW term 
in quantum electrodynamics, then the vector current is 
conserved~\cite{adler1969,bardeen1969}. 
This changes, however, once we consider 
the gauge invariance of the WZW action 
under the full electroweak gauge group of the SM; the  
vector current is no longer conserved. Namely, 
in the presence of spin-one fields which couple to baryon number, 
the baryon vector current is anomalous and new contact interactions 
of pseudo-Chern-Simons form result~\cite{HHH2007,HHH2008}. 
Promoting these spin-one fields to a description of the light vector mesons of QCD, 
$\omega$ and $\rho^{\pm,0}$, Harvey, Hill, and Hill 
recover interactions of the form
$\varepsilon^{\mu\nu\rho\sigma}\omega_\mu Z_\nu F_{\rho \sigma}$  and 
$\varepsilon^{\mu\nu\rho\sigma}\rho^\pm_\mu W^\mp_\nu F_{\rho \sigma}$, e.g., 
where $F_{\rho\sigma}$ is the electromagnetic tensor, and term them
pseudo-Chern-Simons (pCS) interactions~\cite{HHH2007,HHH2008}. 

Such pCS interactions can also be recovered in a chiral Lagrangian framework without
explicit reference to vector mesons. 
In a theory of nucleons and pions with a complete set of 
electroweak gauge fields, treated as external sources, they 
appear at next-to-next-to-leading order (N$^2$LO) in the chiral expansion~\cite{hill2010}. 
To connect to the earlier analysis, the terms 
expressed in  $\omega$ and $\rho$ degrees of 
freedom contribute to certain low-energy constants of the chiral Lagrangian 
and could well ``saturate'' the values of 
these coefficients~\cite{ressat}. However, 
dynamics involving the $\Delta$ resonance also contribute 
and are numerically important in the neutral-current 
sector~\cite{hill2010}. 
At the lowest energies the effective action takes the 
form~\cite{HHH2007}
\begin{equation}
S_{\rm HHH} = \sum_f \xi \int d^4x\, \epsilon^{\mu \kappa \rho\sigma} {\overline N} \gamma_\mu N
{\overline \nu}_L^f \gamma_\kappa {\nu}_L^f F_{\rho\sigma} \,; 
\end{equation} 
we adopt its charged-current analogue for our studies in $\beta$-decay. 
Working in the SM and asserting 
vector-meson dominance one expects 
$\xi \equiv \xi_\omega = ({N_c}/{12 \pi^2})({g_\omega^2}/{m_\omega^2})({e G_F}/\sqrt{2})$~\cite{HHH2007}. 
These pCS interactions could play a role in a swath of processes. 
Possibilities include the interpretation of the low-energy event excess measured by
the MiniBooNE collaboration~\cite{HHH2007,hillmini2010}, 
as well as that of parity-violating observables in reactions such 
as $np \to d \gamma$~\cite{HHH2007,hilltalk}. 
We consider the simplest possible system which can reveal the effect: 
neutron radiative $\beta$-decay~\cite{svgdhe}.

Generally the baryon vector current is anomalous in theories
in which the gauge fields couple differently to $L$- and $R$-handed quarks, 
so that we suppose the contact interactions discussed by 
Harvey, Hill, and Hill possess broader support. 
For example, they ought exist in theories
beyond the SM which possess a SU(2)$_L\times$U(1)$_Y$ gauge symmetry at low energy. 
This notion is not very restrictive given current empirical constraints 
on extended gauge sectors from precision electroweak measurements and 
from flavor physics, note Ref.~\cite{techni} for a review of these in the context
of technicolor models. 

\begin{figure}
\begin{center}
\includegraphics[height=5cm]{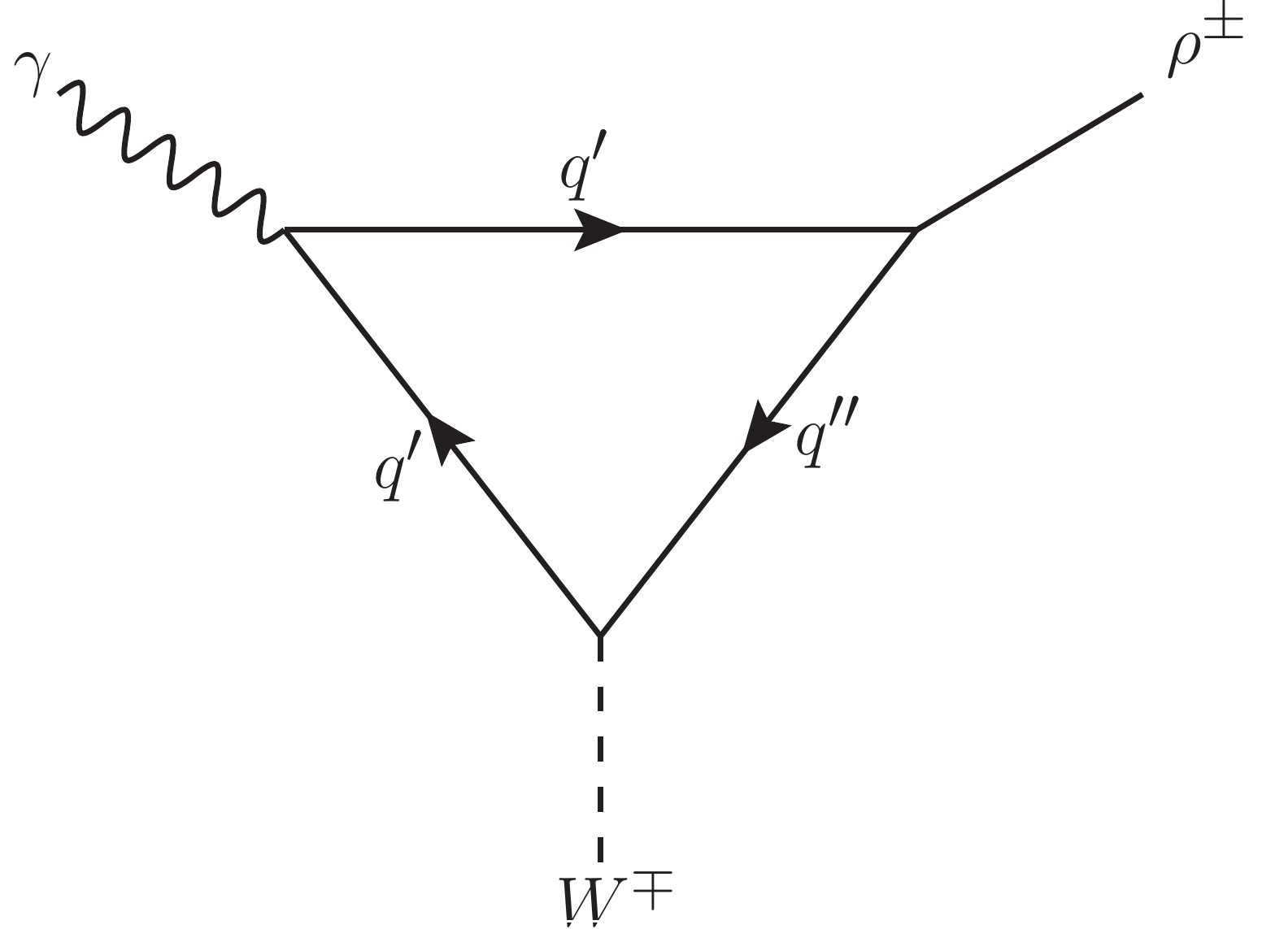}
%\vspace{-0.2in}
\caption{A possible contribution to a pCS interaction at low energy.}
\label{fig:BSM}
\vspace{-0.3in}
\end{center}
\end{figure}

The study of radiative $\beta$-decay can yield direct constraints on $\mathrm{Im}\,\xi$; 
thus we are searching for evidence of new sources of CP violation in 
remnants of the baryon vector current anomaly at low energy. The 
coupling between new matter and the gauge particles can be complex; 
we can then obtain a nontrivial value of 
$\mathrm{Im}\,\xi$ through the interference of the SM contribution with processes such as that 
illustrated in Fig.~\ref{fig:BSM}, where the new particles $q'$ and $q''$ are connected by 
isospin raising or lowering operators. 
The new-matter candidates include, e.g., ``quirks''~\cite{okun,okun2,bjorken,guptaquinn,luty} or 
``dark quarks''~\cite{blennow}. 
Interestingly ``quirk'' models can have novel collider signatures as well~\cite{luty,harnik}.

\section{Isolating a pCS interaction in neutron radiative $\beta$-decay}
\label{isol}

The $\varepsilon^{\mu \nu\rho\sigma}$ structure of the pCS interaction 
suggests its symmetry properties can be used to isolate it. 
In $n(p) \rightarrow p(p') + e^-(l_e) + \overline{\nu}_e(l_\nu) + \gamma(k)$ 
decay the interference of the pCS term with the usual $V-A$ terms in the absolute
square of the transition matrix element yields, to leading order in $\xi$, 
\begin{equation}
|{\cal M}|^2_{\mathrm{pCS}} = - 64 M M' \frac{e^2 G_F^2}{2}
\mathrm{Im}\,\overline{\xi} \frac{E_e}{l_e\cdot k} 
(\mathbf{l}_e\times\mathbf{l}_\nu)\cdot\mathbf{k} + \dots \,,
\end{equation}
where $\bar\xi \equiv \xi/(e G_F/\sqrt{2})$.
The pseudo-T-odd interference term is finite as the photon energy 
$\omega \equiv |\mathrm{k}| \to 0$, so that
its appearance is compatible with Low's theorem~\cite{low} as expected. 
Defining $\eta\equiv (\mathbf{l}_e\times\mathbf{l}_\nu)\cdot\mathbf{k}$ 
we partition the $\beta$-decay phase space into regions of definite sign to
form an asymmetry: 
\begin{equation} 
\mathcal{A}(\omega^{\mathrm{min}}) \equiv 
\frac{{\Gamma_{\eta>0} - \Gamma_{\eta<0}}}{{\Gamma_{\eta>0} + \Gamma_{\eta<0}}} \,,
\end{equation} 
where $\omega^{\mathrm{min}}$ is the lowest detectable photon energy. We recall that
the usual bremstrahlung contribution 
is infrared sensitive~\cite{gaponov,bernard}. 
Working to leading recoil order and to linear order in $\bar\xi$, employing the parameters and
formulae of Ref.~\cite{bernard}, we find 
\begin{equation}
{\cal A}(\omega^{\mathrm{min}}=0.035\,\mathrm{MeV}) = - 2.60 \cdot 10^{-3}
\mathrm{Im}\,\bar\xi (\mathrm{MeV}^{-2}) \,\,,\,\,
\mathrm{Br}(\omega^{\mathrm{min}}=0.035\,\mathrm{MeV}) = 1.83\cdot 10^{-3}\
\end{equation}
and 
\begin{equation}
{\cal A}(\omega^{\mathrm{min}}=0.3\,{\mathrm{MeV}}) = - 1.34 \cdot 10^{-2}\mathrm{Im}\,
\bar\xi (\mathrm{MeV}^{-2}) \,\,,\,\,
\mathrm{Br}(\omega^{\mathrm{min}}=0.3\,\mathrm{MeV}) = 8.62\cdot 10^{-5} \,, 
\end{equation}
where $\omega^{\mathrm{max}}=0.782\,\mathrm{MeV}$. 
In neutron radiative $\beta$-decay 
the branching ratio with increasing photon energy quickly becomes 
phase-space limited. 
Nevertheless, since the number of
neutrons needed to measure a significant effect is controlled by 
${\cal A}^2\,\mathrm{Br}$, the use of photons with energies in excess
of $0.3\,\mathrm{MeV}$ would be more efficacious. 
Of course $\mathrm{Im}\,\overline{\xi}$ in the SM is zero, but
{\em if}  $\mathrm{Im}\, \overline{\xi} \approx \overline{\xi}_{\rho}$ 
the asymmetry would nevertheless be very small indeed. Employing vector-meson dominance 
and adopting $g_{\omega}$ and $g_{\rho}$ from 
a phenomenological analysis of nucleon-nucleon ($NN$)
phase shifts~\cite{elster} yields 
$\overline{\xi}_{\rho} \approx 4.6 \times 10^{-7}$ MeV$^{-2}$ for the charged-current
process, to be compared with 
$\overline{\xi}_{\omega} \approx 4.1 \times 10^{-6}$ MeV$^{-2}$ 
for the neutral-current process. 
To estimate the ability to detect an asymmetry we consider the counting rates
from the neutron radiative $\beta$-decay experiment at NIST~\cite{nico,cooper}. 
The $ep$ double coincidence counting rate in that experiment was $20\,\hbox{s}^{-1}$ for a quoted
neutron flux density of $1.1\times 10^8\,\hbox{cm}^{-2}\hbox{s}^{-1}$~\cite{nicopriv}.
Noting that the total available neutron rate 
at the NG-6 end station at NIST is actually some $20$ times larger~\cite{nicopriv}, then 
in one week of running time with $\omega^{\mathrm{min}}=0.3\,\mathrm{MeV}$ 
one could have a statistical error on the asymmetry of 
$\sim 7\times 10^{-3}$. Both the incident flux and the usable beam size are projected to be 
significantly larger in the planned New Guide Hall~\cite{nicopriv}; the estimated
statistical reach on the asymmetry could then approach $\sim 1 \times 10^{-3}$. 
Although the neutron measurements would 
probe regions of model parameter space which are unconstrained by other data, 
it is prudent to consider the triple-product momentum correlation 
in nuclear radiative $\beta$-decays as well. 
We note in contradistinction that EDM limits impose constraints on chirality-changing 
operators~\cite{posritz}. 
The asymmetries will generally be larger in nuclear decays 
because the asymmetry grows with the energy released in the decay. 
In addition, since the pCS interaction engenders parity violation but does not 
involve the nucleon spin, it is coherent at the amplitude 
level~\cite{hill2010}. 
There are different possibilities for its study in nuclei. 
As an example, in $^{19}$Ne decay 
the energy release is some $\sim 2.7$~MeV, and the ${}^{19}$Ne lifetime is $17.2$ s. 
Perhaps one could study $^{19}$Ne $\to$ $^{19}$F radiative $\beta$-decay 
in an atom trap experiment, note Ref.~\cite{trapneon} for a discussion of related work. 
An accelerator experiment might also allow access to the triple product momentum
correlation; the beam momentum could define one of the needed momenta, so that 
one could access the correlation via a forward-backward asymmetry. 
The nuclear studies could possibly be realized at a rare isotope accelerator such as 
the Facility for Rare Isotope Beams (FRIB) at Michigan State University. 
Irrespective of these experimental choices, 
the best candidate nuclear radiative $\beta$-decay
would have a large energy release and be mediated by a pure Fermi transition. 

\section{Induced Correlations from Known FSI} 
\label{induce}

\begin{figure}
\begin{center}
\includegraphics[height=5cm]{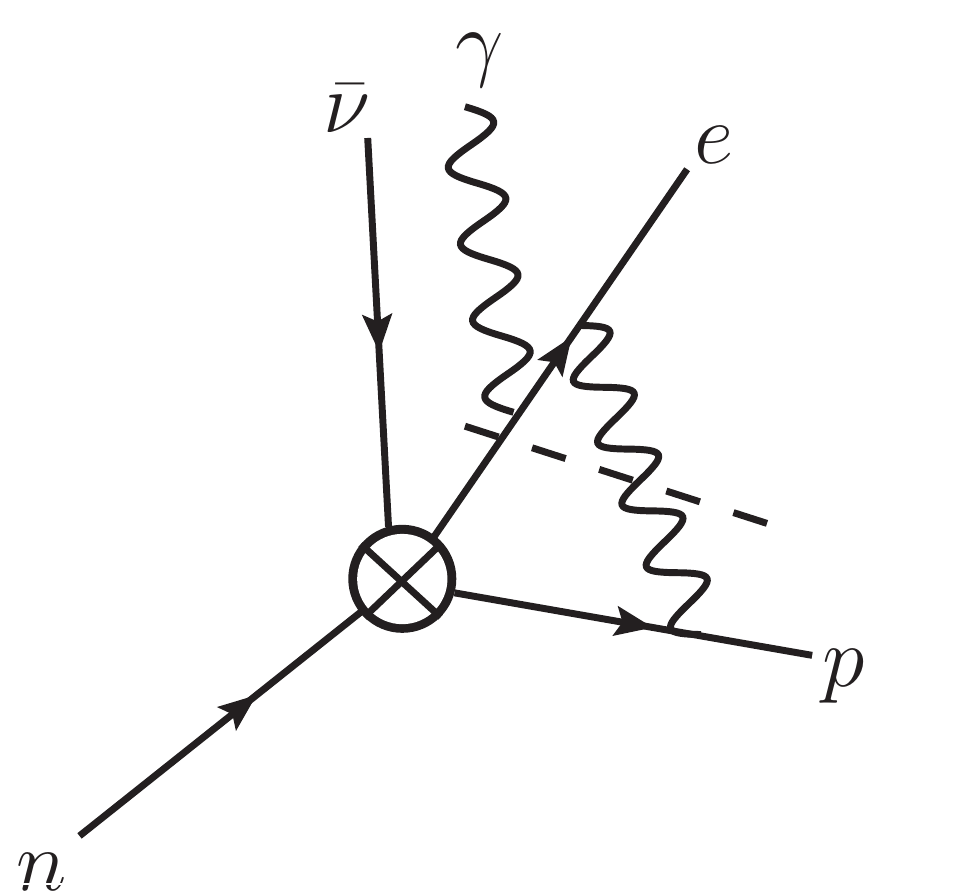}
%\vspace{-0.2in}
\caption{One contribution to the discontinuity across the $\gamma e$ cut.}
\label{fig:cut}
\vspace{-0.3in}
\end{center}
\end{figure}

As long known, ${\cal O}(\alpha)$ 
radiative corrections can mimic pseudo-T-odd correlations~\cite{callantreiman,okunkhrip}
in particle decay and production processes. In neutron radiative $\beta$-decay
such effects are operative as well. A similar calculation has been performed
in kaon radiative $\beta$-decay~\cite{braguta,kubis,khrip2}. 
In neutron radiative $\beta$-decay there are 
three basic types of physical cuts, and the total number of cut diagrams is 14. 
One particular diagram associated with the $\gamma e$ cut 
is shown in Fig.~\ref{fig:cut}.
Using the Cutkosky rules~\cite{cutkosky}, 
we can compute the discontinuity across each physical cut. 
Namely, for the $\gamma e$ cut, we have 
\begin{equation}
-i \hbox{Disc} {\cal M}^{\gamma e} (n\to p e \bar \nu) = 
\sum_f \int \frac{d^4 l_1 d^4 l_2}{(2\pi)^2} \delta^{(4)}(l_1 + l_2 - l_e - k) 
\delta^{(+)}(l_1^2 - m^2) \delta^{(+)}(l_2^2) 
{\cal M}^*(\gamma e \to \gamma e) {\cal M}(n \to p {\bar\nu} e \gamma) \,,
\end{equation}
where $\delta^{(+)}(l^2 - m^2)$ means that a particle of momentum $l$ and mass $m$
has been put on its positive energy shell. We have determined that the 
first nontrivial contribution to the triple momentum correlation is 
in radiative recoil order,
which is encouraging. A complete calculation of the effect induced by 
the ${\cal O}(\alpha)$ electromagnetic correction is in progress~\cite{svgdhe}. 

\section{Other Processes: Radiative Muon Capture}
\label{other}

Both radiative (RMC) and ordinary muon capture (OMC) on the proton can be used
to study the pseudoscalar coupling $g_p$ of the nucleon. The value of
$g_p$ is also predicted by 
the partially conserved axial current hypothesis (PCAC), which fits
under the aegis of modern (heavy-baryon) chiral 
perturbation theory (HBChPT)~\cite{kaiser,kaiser2}. 
After much controversy, there has been significant experimental progress: the 
value of $g_p$ recently determined by the MuCap collaboration 
agrees with the HBChPT prediction~\cite{mucap,winter}. Nevertheless, disagreements 
of the old OMC and RMC results with theory and with each other persist~\cite{winter}. 
We note that the pCS interaction
proposed by Harvey, Hill, and Hill in the SM would contribute exclusively to RMC, so that if 
is of the proper numerical size and if 
the ortho-para transition rate in muonic molecular hydrogen 
is described well by theory, making the old OMC result consistent with MuCap, 
then all the residual discrepancies could be reasonably resolved. 
The numerical size of the pCS term in 
this context is under evaluation~\cite{svgdhe2}. 
A plurality of effects could influence the value of $g_P$ extracted from RMC; notably 
contributions from $\Delta$ resonance degrees of freedom have been evaluated
and are too small to explain the discrepancy~\cite{bernardrmc}.

\section{Summary}
\label{conc}

Harvey, Hill, and Hill suggest that remnants of the
baryon vector current anomaly in the Standard Model exist in low-energy 
interactions~\cite{HHH2007,HHH2008}. 
We have argued that the new contact interactions they discuss 
can also appear in theories beyond the SM with SU(2)$_L \times$U(1)$_Y$ electroweak
symmetry at low energies. We have studied how new sources of CP violation connected to such 
interactions can be probed through a triple-product momentum correlation
in neutron radiative $\beta$-decay. 
The pCS interaction does not involve the nucleon spin; the
constraints offered through the study of the pseudo-T-odd,
P-odd asymmetry in neutron (or nuclear) radiative $\beta$-decay 
are complementary but distinct from those from EDMs. 



%% References


%\bibliographystyle{elsarticle-num}
%\bibliography{trace}



\end{document}